\documentclass{pasj00}

\SetRunningHead{Akiyama et al.}{Multi-epoch VERA observations of Sgr A*. I}

\Received{2012/05/22}
\Accepted{2013/05/13}
\Published{2013/08/25}

\newcommand{\jcap}{JCAP}

\newcommand{\uas}{$\rm \mu$as }

\newcommand{\mdot}{\dot{m}}

\begin{document}

%-------------------
% Title
%-------------------
\title{Multi-Epoch VERA Observations of Sagittarius A*:\\I. Images and Structural Variability}

%-------------------
% Author
%-------------------
\author{Kazunori Akiyama,\altaffilmark{1,2,*} Rohta Takahashi,\altaffilmark{3} Mareki Honma,\altaffilmark{2,4} Tomoaki Oyama,\altaffilmark{2} \and Hideyuki Kobayashi\altaffilmark{1,2}}

%-------------------
% Affiliation
%-------------------
\affil{\altaffilmark{1} Department of Astronomy, Graduate School of Science, The University of Tokyo, 7-3-1 Hongo, Bunkyo-ku, Tokyo 113-0033, Japan}
\affil{\altaffilmark{2} Mizusawa VLBI Observatory, National Astronomical Observatory of Japan, 2-21-1 Osawa, Mitaka, Tokyo 181-8588}
\affil{\altaffilmark{3} Department of Natural and Physical Sciences Tomakomai National College of Technology, Tomakomai 059-1257, Japan}
\affil{\altaffilmark{4} Graduate University for Advanced Studies, Mitaka, 2-21-1 Osawa, Mitaka, Tokyo 181-8588}

%-------------------
% Keywords
%-------------------

%-------------------
% Maketitle ! Do not change!
%-------------------
\maketitle

%-------------------
% Abstract
%-------------------
\begin{abstract}
We report the results of multi-epoch observations of Sgr A* with VLBI Exploration of Radio Astrometry (VERA) at 43 GHz, carried out from 2004 to 2008. We detected a time variation of flux at 11 \% level and intrinsic size at 19 \%. In addition, comparisons with previous Very Long Baseline Array (VLBA) results shows that Sgr A* underwent the flaring event at least longer than 10 days in May 2007. The intrinsic size of Sgr A* remained unchanged within 1 $\sigma$ level from the size before/after the flaring event, indicating that the brightness temperature of Sgr A* was increased. The flaring event occurred within 31 d, which is shorter than the refractive time scale. Moreover it is difficult to explain the increase in the spectral index at the flaring event by the simple interstellar scattering model. Hence, the flaring event is most likely associated with the changes in intrinsic properties of Sgr A*. We considered the origin of the brightness temperature variation, and concluded that the flaring event of Sgr A* could be explained by the continuous heating of electrons, such as a standing shock in accretion flow.
\end{abstract}

%
%---------------------------------------------------------------------------------------
% Section Introduction
%---------------------------------------------------------------------------------------
\section{Introduction\label{sec:Introduction}}
\renewcommand{\thefootnote}{\fnsymbol{footnote}}
\footnotetext[1]{Research Fellow of the Japan Society for the Promotion of Science}
\renewcommand{\thefootnote}{\arabic{footnote}}
%Sgr A* is massive blackhole %
There is plenty of evidence that the center of our galaxy hosts a massive black hole with a mass of approximately 4 $\times$ 10$^6$ M$_\odot$ at a distance of about 8 kpc. The compact radio source Sagittarius A* (Sgr A*), located in the galactic center is believed to be associated with this black hole. The measurements of the proper motion of Sgr A* showed that Sgr A* must contain $\verb|>|$ 4 $\times$ 10$^5$ M$_\odot$ \citep{Reid2004} and provided support for the existence of a massive black hole in the vicinity of Sgr A*.

%Emission Mechanism%
The emission from Sgr A* is detected at the radio, infra-red and X-ray band. The bolometric luminosity of Sgr A*, which is $\sim$ $10^{36}$ erg s$^{-1}$, is $\sim$ 8.5 orders of magnitude smaller than the Eddington limit for its black hole mass of $\sim$ 4 $\times$ 10$^6$ M$_\odot$ in spite of the existence of the large gas reservoir from stellar winds in its vicinity. A number of theoretical models have been proposed to explain the low luminosity of Sgr A*. One of the successful models that can explain the observed spectrum is Radiatively Inefficient Accretion Flow (RIAF), which has a low radiative efficiency \citep{Narayan1998,Yuan2003,Gammie2009,Kato2009,Moscibrodzka2009,Yuan2009,Dexter2010}. One of the most improved RIAF models that explains the observed spectrum phenomenologically is a solution involving substantial mass loss (although the outflowing mass is ignored in calculating the radio emission) and a non-thermal component of electrons \citep{Yuan2003}. In addition to the models which explain the emission of Sgr A* in terms of the accretion flow, there exist other models in which the emission of Sgr A* is mainly attributed to an outflow such as sub-relativistic or relativistic jet \citep{Falcke2000,Yuan2002}. Another model with escaping wind consisting of thermal electrons is also proposed \citep{Loeb2007}.

%Scattering%
The apparent size of the Schwarzschild radius of the black hole is about 10 \uas and is the largest among black hole candidates. Thus, the spatial resolution of VLBI can resolve the structure in the vicinity of the Sgr A*. However, the direct imaging of the intrinsic structure of Sgr A* is still difficult at the radio band due to the interstellar scattering \citep{Narayan1989a,Narayan1989b}. Measurements of the apparent size of Sgr A* showed that the size is proportional to $\lambda^2$ at a wavelength longer than $\sim$ 1.3 cm, while the intrinsic structure appears to be detectable at a wavelength shorter than $\sim$ 1.3 cm \citep{Lo1998,Bower2004,Bower2006,Shen2005,Falcke2009}. The frequency-dependency of the intrinsic size of Sgr A* in the radio bands \citep{Shen2005,Bower2006} suggests that the intrinsic size of Sgr A* at radio band is determined by the photosphere of optically thick plasma \citep{Loeb2007,Falcke2009}. Based on the existence of the sub-mm bump in the broadband spectrum, the transition between the optically thin and optically thick regimes is expected to occur in the sub-mm band.

%Intraday variation%
Sgr A* is known to have variabilities at radio, IR and X-ray band with time-scales ranging from intra-day to a few months. The intra-day variation of Sgr A* has been studied intensively by international multi-wavelength campaigns \citep{Yusef2008,Yusef2009}. The radio band variability on time scales longer than a day has been discussed mainly based on observations with connected interferometers \citep{Macquart2006}. As an origin of the variation, both intrinsic and extrinsic models are proposed, but the mechanism of the time variation is still under discussion. Interstellar scintillation is the primary mechanism that may cause extrinsic variability. Previous studies suggested that if the scattering medium of Sgr A* lies on a thin screen, all the observed flux variability must be intrinsic to the source itself \citep{Macquart2006}. Meanwhile, an extended scattering region may explain the broad characteristics of the variability longer than 4 days \citep{Macquart2006}. However, the physical structures associated with this extended scattering medium is unknown. 

Recent studies at cm/mm wavelengths reported the existence of long timescale flux variations at short, cm/mm wavelengths where interstellar effects are negligible. A massive monitoring of Sgr A* with VLA and the detections of high fluxes of Sgr A* at mm wavelengths indicate that the flux variation at mm wavelengths is likely to be not explainable by a simple model of interstellar scattering \citet{Rickett1990} but intrinsic \citep{Herrnstein2004,Lu2011}. In addition, recent Event Horizon Telescope observations discovered that flux density of Sgr A* at 1.3 mm had a day-to-day variability without a variation in its size indicating a variation in a brightness temperature in innermost region of Sgr A* \citep{Fish2011}.
%, and also indicate a possible bimodal distribution of flux density such as low/high-flux density state at these frequencies may exist \citep{Herrnstein2004,Lu2011}. In addition, recent Event Horizon Telescope observations discovered the transition of the steady state, though it is unclear whether this transition is related to the transition between low and high-flux density state \citep{Fish2011}.

% VLBI study and Motivation of this study %
High spatial resolution of VLBI enables monitoring of the flux and structure in the vicinity of a black hole.
Hence, a monitoring observation with VLBI is of great importance to investigate a mechanism of time-variation of Sgr A*.
However, there have been no VLBI studies aiming to monitor the time variation of both the flux and structure of Sgr A* in time scales longer than a month.
Previous VLBI studies on the size of Sgr A* found that the intrinsic structure can be estimated from VLBI observations at a frequency higher than $\sim$ 22 GHz.
Moreover, simultaneous observations of Very Long Baseline Array (VLBA) and Very Large Array (VLA) indicated that the total flux of Sgr A* obtained by a connected array tend to be over-estimated by sampling the diffuse emission from the vicinity of the Sgr A* \citep{Yusef2009}.
Thus, VLBI observations are most suitable for a detailed study of the long-term variation of the Sgr A*.
Hence, in order to investigate the relation between the flux and intrinsic size of Sgr A*, we have carried out the observations of Sgr A* from 2004 to 2008 with VLBI Exploration of Radio Astrometry (VERA) at 43 GHz, and here we report the results.

%about this paper%
The plan of this paper is as follows: in \S 2, we describe the observations and the data reduction.
In \S 3, we present observed images (\S 3.1.1), fluxes and structures obtained from model-fitting (\S 3.1.2 and \S 3.2).
In \S 4, we discuss the observed time variation of the brightness temperature of Sgr A*.
Finally, we summarize the results and discussions in \S 5.
Throughout this paper, the mass of Sgr A* $M$ is assumed to be $4 \times 10^6$ M$_\odot$, and the distance to Sgr A* $D$ is assumed to be 8 kpc. 

%---------------------------------------------------------------------------------------
% Section Observations amd Reductions
%---------------------------------------------------------------------------------------
\section{Observations and Reductions\label{sec:Observations and Reductions}}
\subsection{Observations\label{subsec:Observations}}
% General introduction %
VERA observations of Sgr A* at 43 GHz were regularly performed between November, 2004 and April, 2009.
The observations were done in the dual-beam mode observing Sgr A* and J1745-2820 simultaneously for aiming at astrometry of Sgr A*.
NRAO 530 is also observed in several scans for checking the consistency of the amplitude calibration.
In this paper, we focus only on the data of Sgr A*.
Results of the astrometry of Sgr A* will be discussed elsewhere.

% Observing epoch %
In some epochs, one or more stations were partly or fully missed due to system trouble or/and bad weather.
Since the VERA array consists of only four stations, a lack of stations causes severe degrading of synthesized images, and would introduce a large error in quantities derived from images or model-fitting toward visibilities.
Hence, in this paper, we use the data of 10 epochs for which full stations are available under relatively good conditions.
The epochs presented here are: day of year (DOY) 294 in 2005, 079, 109 and 308 in 2006, 073, 093 and 264 of  2007, 076, 085 and 310 of  2008 (October 21 of  2005, March 20, April 19 and November 4 of  2006, March 14, April 3 and September 21 of  2007, March 16, March 25, November 8 of 2008). 
The system noise temperatures at the zenith were typically 400-600 K through all the epochs.

% Instrumental setting %
Left-hand circular polarization (LHCP) signals were received and sampled with 2-bit quantization, and filtered using the VERA digital filter unit \citep{Iguchi2005}.
The data were recorded at a rate of 1024 Mbps, providing a bandwidth of 256 MHz.
One of two IF-channels of 128 MHz bandwidth was assigned to Sgr A* (and also NRAO 530).
Correlation processes were performed with the Mitaka FX correlation \citep{Chikada1991}.

\subsection{Data Reductions\label{subsec:Data Reductions}}
% Figure vera_uvplot%
\begin{figure}[!h]
\begin{center}
%\showthe\columnwidth
\includegraphics[width=0.8\hsize]{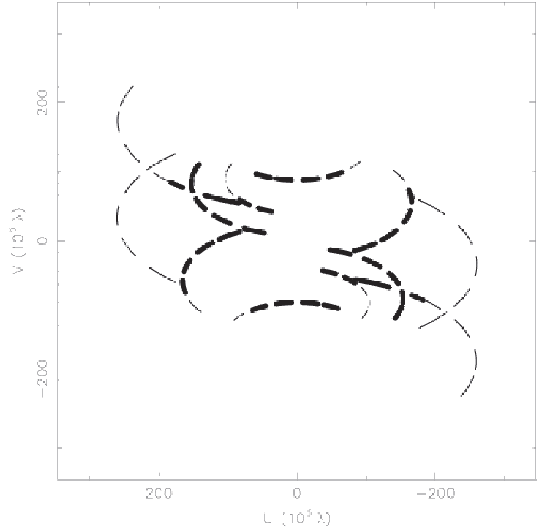}
\caption{The distribution of visibilities on the UV plane of epoch (j).
Thin lines indicate all visibilities sampled during observation, while bold lines indicate UV for which fringes are detected.
All detections are concentrated in baselines shorter than 200 M$\lambda$.
\label{fig:vera_uvplot}}
\end{center}
\end{figure}

% Figure sgra_projplot_r08310a%
\begin{figure}[!h]
\begin{center}
%\showthe\columnwidth
\includegraphics[width=0.8\hsize]{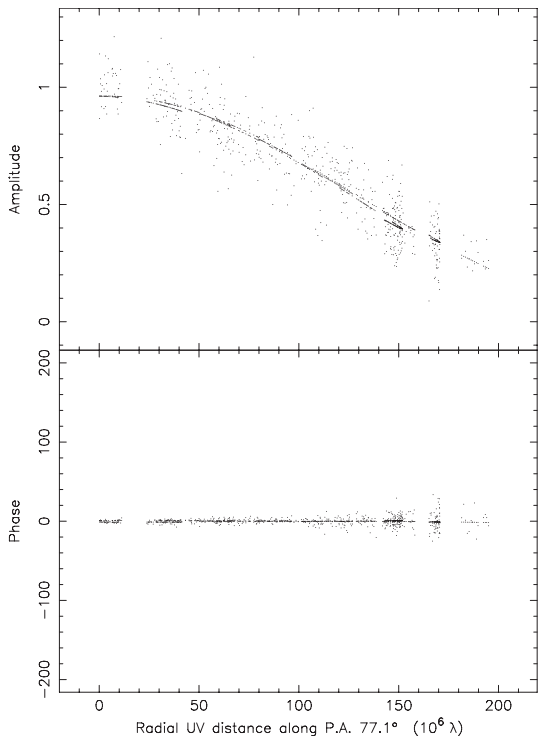}
\caption{The visibility amplitude and phase of epoch (j) as a function of projected uv-distance for Sgr A*. All visibilities are plotted along a PA of 77.1 degrees East of North, corresponding to the major axis of the Gaussian model of epoch (j). The solid lines indicate the model obtained by elliptical Gaussian fitting. Parameters of the Gaussian model are shown in Table \ref{tab:UVFIT results}.
\label{fig:sgra_projplot_r08310a}}
\end{center}
\end{figure}

% Figure sgra_cpplot_r08310a%
\begin{figure}[!h]
\begin{center}
%\showthe\columnwidth
\includegraphics[width=0.8\hsize]{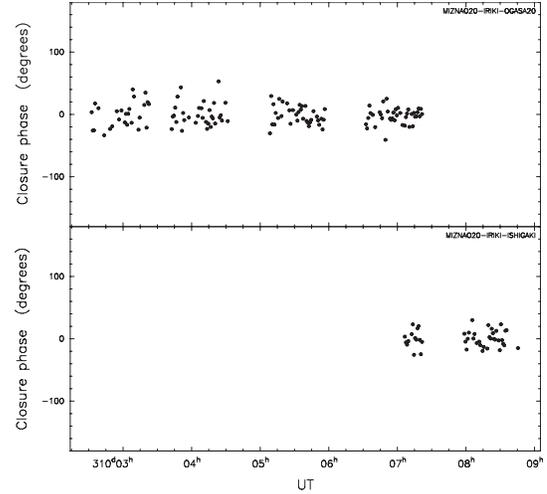}
\caption{The closure phase of epoch (j) as a function of Universal Time.
\label{fig:sgra_cpplot_r08310a}}
\end{center}
\end{figure}

% AIPS Reduction % 
Data reductions were carried out with the NRAO Astronomical Imaging Processing System (AIPS) first, using the standard algorithms including phase and delay calibrations and fringe-fitting.
A standard {\it a priori} amplitude calibration was performed using the AIPS task APCAL based on the measurements of system temperature ($T_{sys}*$), which were obtained based on the chopper-wheel method during observations.
Delays and rates were determined with fringe-fitting directly to the Sgr A* visibilities using the AIPS task FRING. We set a solution interval of $\sim$ 1 min and minimum S/N ratio of = 3.
To avoid false detections, we set a practical search window of 20 nsec $\times$ 50 mHz. In this search window, the false detection probability is $\sim$ 5 \%.
Since Sgr A* has still scattered broad structure at 43 GHz and its FWHM size is $\sim$ 0.7 mas at 43 GHz, fringes were not detected for most visibilities observed in baselines longer than 200 M$\lambda$.
Figure \ref{fig:vera_uvplot} shows a typical uv-coverage of Sgr A*.
In fact, all the detected fringes are located within 200 M$\lambda$.
Visibilities related to the closure of Mizusawa, Iriki and Ogasawara stations are mostly valid, while more than 70 $\%$ of visibilities sampled with Ishigaki-jima station are invalid, since Ishigaki-jima station mainly provides baselines exceeding 200 M$\lambda$.
As an example of the visibilities after applying all calibrations, we show the uv-distance plot of the data obtained in November 8 of 2008 (in later sections, named as epoch (j)) in Figure \ref{fig:sgra_projplot_r08310a}.
The horizontal axis of Figure \ref{fig:sgra_projplot_r08310a} is the uv-distance projected along the major axis of the elliptical Gaussian models of visibilities (see, Section 3.2).
The top panel of Figure \ref{fig:sgra_projplot_r08310a} shows that the visibility amplitude decreases with uv-distance down to $\sim$ 200 mJy at 200 M$\lambda$.
Moreover, from the distribution of the visibility phases (the bottom panel), one can see that the scatter of phases becomes larger for longer baselines, which is due to lower S/N ratio for longer baselines.
The minimum detected flux in Figure \ref{fig:sgra_projplot_r08310a}, $\sim$ 200 mJy, is comparable to the fringe detection limit of VERA under typical conditions.
In fact, the minimum detectable flux is $\sim$ 200 mJy for S/N = 3 and $T_{\rm sys}*=500$ K).
This fact is consistent with the non-detection of fringe-fitting solutions for baselines longer than 200 M$\lambda$.

% Difmap Imaging %
After the fringe fitting, calibrated data were output to DIFMAP.
We averaged visibilities to $\sim$ 60 s in DIFMAP.
We checked the closure phase of Sgr A* and confirmed that the closure phase of Sgr A* is obtained throughout almost all the time in each epoch, and that their S/N ratios are high enough to apply self-calibration to the visibility phases.
As an example, we show the closure phases of epoch (j) in Figure \ref{fig:sgra_cpplot_r08310a}.
As seen in Figure \ref{fig:sgra_cpplot_r08310a}, closure phases of the combination of Mizusawa, Iriki and Ogaswara stations are mostly determined, while more than 60 $\%$ of the closure phases of the combination of Mizusawa, Ogasawara and Ishigaki stations are missed because the sources are mostly resolved out with long baselines.
Two other closures including Ishigaki-jima station were totally invalid.
Finally, we carried out imaging of Sgr A* by iterating the self-calibration of visibility phase and CLEAN using DIFMAP. 

\subsection{Model Fitting\label{subsec:Model Fitting}}

% Gaussian fitting %
After imaging, we also conducted Gaussian fitting to the calibrated visibilities, since most previous VLBI studies of Sgr A* also adopted this technique.
We output visibilities to text files using AIPS task PRTUV, and fitted Gaussian models to visibilities directly by using the least-square method.
We confirmed that the results of Gaussian fitting, which is obtained with our own code of least-square fit, coincides with 
the results obtained with the Gaussian fitting program in DIFMAP. 

%Bootstrap Method%
For estimating the fitting errors of Gaussian parameters, we used the non-parametric Percentile Bootstrap method \citep{Wall2003}.
This method is straightforward in deriving estimates of confidence intervals of fitted parameters.
We briefly summarize a process of estimating confidence intervals: at first, we created a data-set which has the same number of data by re-sampling the observed visibilities allowing repetition of the data.
Re-sampling was simply done using a uniform random-number generator.
Then, we fitted the Gaussian model to the new data-set and obtained a bootstrap estimate.
We repeated this process 30,000 times and obtained 30,000 sets of the bootstrap estimates.
The Percentile bootstrap confidence limits of each parameter are obtained as the edges of the middle 99.7 $\%$ fractions of the Bootstrap estimates.
We confirmed that the averages of bootstrap parameters coincides to the results of the least-square method within 1 $\%$.
Finally, the errors of each parameter are obtained as one-third of displacements between the averages of bootstrap parameters and the confidence limits, so that the obtained error corresponds to 1-$\sigma$ uncertainty.
The obtained errors are slightly greater than the standard errors estimated from the co-variance matrix of the least-square method.
This means that the bootstrap method provides more conservative error estimates than the standard errors of the least-square method.

% Systematic Error %
The systematic errors of the total flux are estimated as below. In our observations, it is difficult to estimates systematic errors using calibrators, since our observations were carried out over an interval longer than one month and calibrators
are also variable in such time scales. 

Instead, we estimated 10 \% of the total flux as a systematic error, and added them in quadrature to the fitting error. This is
a reasonable estimate, since the amplitude calibration was done by a-priori calibration using the system temperature $T_{sys}*$ measured by the chopper-wheel method.

% Figure NRAO530_flux%
\begin{figure}
\begin{center}
%\showthe\columnwidth
\includegraphics[width=0.8\hsize]{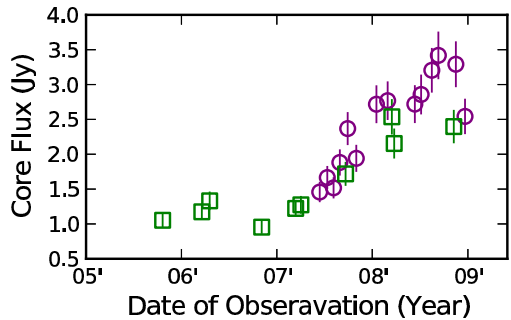}
\caption{The time variations of the core-flux of NRAO 530. The squares (colored green in the on-line version) indicate our results, while the circles (colored purple in the on-line version) indicate VLBA results.
\label{fig:NRAO530_flux}}
\end{center}
\end{figure}

For checking the fairness of amplitude calibration, we also analyzed NRAO 530 which was observed simultaneously as an amplitude calibrator.
We confirmed that the r.m.s. of gain solutions obtained from the amplitude self-calibration of NRAO 530 were around or less than 10 $\%$ for most epochs.
Figure \ref{fig:NRAO530_flux} shows the core-flux variations of NRAO 530.
These core-fluxes were obtained by the elliptical Gaussian-fitting to the visibilities and their errors were estimated in the same way as Sgr A*.
In addition to VERA data, Figure \ref{fig:NRAO530_flux} shows the core-fluxes obtained from data of the Boston University blazar monitoring program with the VLBA\footnote{http://www.bu.edu/blazars/VLBAproject.html}.
These data were obtained by model-fitting to calibrated visibilities.
Figure \ref{fig:NRAO530_flux} suggests the trend of core-flux variation obtained from VERA and VLBI is presumably consistent, though both observations were not carried out simultaneously.

\subsection{Calculation of intrinsic size\label{subsec:Calculation}}
After model fitting, we calculated the intrinsic size of Sgr A* using the total flux and the major and minor axes sizes obtained from Gaussian-fitting.
We estimated the major axis intrinsic size based on a traditional method used in previous VLBA observations \citep{Lo1998,Bower2004,Bower2006,Falcke2009}.
We adopted the scattering law of \citet{Falcke2009}, in which the scattering size $\phi_{\rm scatt}$ was given by
\begin{equation}\label{eq:major scatter}
 \phi_{\rm scatt}=1.36 {\rm \,mas}\, \times \left( \frac{\lambda}{\rm 1 \, cm} \right) ^2,
\end{equation}
yielding $\phi_{\rm scatt}$ of $\sim$ 661 \uas at 43.0 GHz.
Assuming that the intrinsic brightness distribution of Sgr A* is elliptical Gaussian, %using $\phi_{\rm scatt}$ and the observed size of the major axis $\phi_{\rm maj}$, 
the intrinsic size $\phi_{\rm int}$ is given by
\begin{equation}
\phi_{\rm int}=\sqrt{\phi_{\rm maj}^2-\phi_{\rm scatt}^2}.
\end{equation}
We note that here the scattering effect is assumed to be the same throughout all the epochs.
To convert units of intrinsic size $\phi _{\rm int}$, one can refer to following relation between angular scale in \uas and Schwarzchild radius ${\rm R_s}$ given by
\begin{equation}\label{eq:schwarzchild}
1\,{\rm R_s}=9.87 {\rm\,\mu as}\, \times \left( \frac{D}{\rm 8 \,kpc} \right) ^{-1} \left( \frac{M}{4 \times 10^6 \,M_\odot}\right).
\end{equation}

\subsection{VLBA data of Sgr A*\label{subsec:VLBA data of Sgr A*}}
% Table VLBA results%
\begin{table*}
\begin{center}
\caption{The source structure derived from previous VLBA observations.\label{tab:VLBA results}}
\begin{tabular}{ccccccc}\hline \hline
Epoch &Freq.&$S_{\rm total}$&$\phi _{\rm maj}$&$\phi _{\rm int}$&$R _{\rm int}$&Reference and comment\\ 
(yyyy-mm-dd)&(GHz)&(Jy) &($\rm \mu$as)&($\rm \mu$as) &(R$_{\rm g}$) & \\ \hline
1994-04-26&43.151&&720$^{+10}_{-10}$&296$^{+24}_{-24}$&30.0$^{+2.5}_{-2.5}$&\citet{Shen2005}\\
1994-09-29&43.151&1.28$^{+0.16}_{-0.16}$&762$^{+38}_{-38}$&387$^{+75}_{-75}$&39.2$^{+7.6}_{-7.6}$&\citet{Bower1998}\\
1994-09-29&43.151&&720$^{+10}_{-10}$&296$^{+24}_{-24}$&30.0$^{+2.5}_{-2.5}$&\citet{Shen2005}\\
1997-02-01&43.2&1.03$^{+0.10}_{-0.10}$&700$^{+10}_{-10}$&247$^{+28}_{-28}$&25.0$^{+2.9}_{-2.9}$&\citet{Lo1998}\\
1997-02-14&43.213&&728$^{+16}_{-11}$&319$^{+12}_{-8}$&32.3$^{+1.2}_{-0.8}$&\citet{Bower2004}\\
1997-02-14&43.213&&710$^{+10}_{-10}$&275$^{+26}_{-26}$&27.9$^{+2.6}_{-2.6}$&\citet{Shen2005}\\
1999-03-23&43.135&&710$^{+10}_{-10}$&269$^{+26}_{-26}$&27.3$^{+2.7}_{-2.7}$&\citet{Shen2005}\\
1999-04-24&43.135&&690$^{+10}_{-10}$&211$^{+33}_{-33}$&21.4$^{+3.3}_{-3.3}$&\citet{Shen2005}\\
1999-05-23&43.1&&713$^{+12}_{-9}$&275$^{+11}_{-8}$&27.8$^{+1.1}_{-0.8}$&\citet{Bower2004}\\
2001-07-12&43.2&&725$^{+22}_{-12}$&311$^{+17}_{-9}$&31.5$^{+1.7}_{-0.9}$&\citet{Bower2004}\\
2001-07-29&43.2&&770$^{+30}_{-18}$&405$^{+19}_{-12}$&41.0$^{+1.9}_{-1.2}$&\citet{Bower2004}\\
2001-08-05&43.2&&704$^{+64}_{-43}$&258$^{+58}_{-39}$&26.2$^{+5.9}_{-4.0}$&\citet{Bower2004}\\
2002-04-15&43.2&&708$^{+17}_{-13}$&269$^{+15}_{-11}$&27.2$^{+1.5}_{-1.1}$&\citet{Bower2004}\\
2002-05-03&43.2&&708$^{+6}_{-4}$&269$^{+5}_{-4}$&27.2$^{+0.5}_{-0.4}$&\citet{Bower2004}\\
2002-05-13&43.2&&709$^{+9}_{-6}$&272$^{+8}_{-6}$&27.5$^{+0.8}_{-0.6}$&\citet{Bower2004}\\
2004-03-08&43.175&&722$^{+2}_{-2}$&302$^{+5}_{-5}$&30.6$^{+0.5}_{-0.5}$&\citet{Shen2005}\\
2004-03-20&43.175&&725$^{+2}_{-2}$&309$^{+5}_{-5}$&31.3$^{+0.5}_{-0.5}$&\citet{Shen2005}\\
2007-05-15&43.1&2.02$^{+0.22}_{-0.22}$&710$^{+10}_{-10}$&267$^{+27}_{-27}$&27.0$^{+2.7}_{-2.7}$&\citet{Lu2011}\\
2007-05-16&43.1&1.59$^{+0.17}_{-0.17}$&720$^{+10}_{-10}$&292$^{+25}_{-25}$&29.6$^{+2.5}_{-2.5}$&\citet{Lu2011}\\
2007-05-17&43.1&1.99$^{+0.21}_{-0.21}$&720$^{+10}_{-10}$&292$^{+25}_{-25}$&29.6$^{+2.5}_{-2.5}$&\citet{Lu2011}\\
2007-05-18&43.1&1.61$^{+0.17}_{-0.17}$&710$^{+10}_{-10}$&267$^{+27}_{-27}$&27.0$^{+2.7}_{-2.7}$&\citet{Lu2011}\\
2007-05-19&43.1&1.86$^{+0.20}_{-0.20}$&710$^{+10}_{-10}$&267$^{+27}_{-27}$&27.0$^{+2.7}_{-2.7}$&\citet{Lu2011}\\
2007-05-20&43.1&1.66$^{+0.18}_{-0.18}$&720$^{+10}_{-10}$&292$^{+25}_{-25}$&29.6$^{+2.5}_{-2.5}$&\citet{Lu2011}\\
2007-05-21&43.1&2.02$^{+0.22}_{-0.22}$&720$^{+10}_{-10}$&292$^{+25}_{-25}$&29.6$^{+2.5}_{-2.5}$&\citet{Lu2011}\\
2007-05-22&43.1&1.90$^{+0.20}_{-0.20}$&720$^{+10}_{-10}$&292$^{+25}_{-25}$&29.6$^{+2.5}_{-2.5}$&\citet{Lu2011}\\
2007-05-23&43.1&1.92$^{+0.21}_{-0.21}$&720$^{+10}_{-10}$&292$^{+25}_{-25}$&29.6$^{+2.5}_{-2.5}$&\citet{Lu2011}\\
2007-05-24&43.1&1.78$^{+0.19}_{-0.19}$&680$^{+10}_{-10}$&172$^{+40}_{-40}$&17.4$^{+4.0}_{-4.0}$&\citet{Lu2011}\\
2007-05-15$\sim$24&43.1&1.79$^{+0.19}_{-0.19}$&710$^{+10}_{-10}$&267$^{+27}_{-27}$&27.0$^{+2.7}_{-2.7}$&average of \citet{Lu2011}\\ \hline
\end{tabular}
\end{center}
\end{table*}

To trace the variation of Sgr A* better, we combined the results of previous VLBA observations with our results.
We calculated the major axis intrinsic size from data in the same way as \S2.4.
We show the data in Table \ref{tab:VLBA results}.
These data are taken from the following references: \citet{Bower1998,Lo1998,Bower2004,Shen2005,Lu2011}.
\citet{Bower2004} and \citet{Shen2005} used the closure-amplitude method for fitting Gaussian-models to visibilities.
The use of closure amplitude discards the information of total flux.
Hence, the only available parameters are the sizes of major and minor axis.
\citet{Bower1998}, \citet{Lo1998} and \citet{Lu2011} used visibility fitting.
Thus, parameters of both the sizes and total flux are available.
To consider the systematic error of {\it a priori} amplitude calibration, we add 10 $\%$ of the total flux to the error originally reported.
%{\bf [deleted: About the data of \citet{Lu2011}, we adopted time-averaged values of all 10 epochs].}d
In Table 3, we also present intrinsic sizes that are calculated in the same manner as those in \S3.2. 
We note that observed on September 9 of 1994 and February 14 are obtained by the same observation, but analyzed by different methods \citep{Bower1998,Bower2004,Shen2005}. We adopt all the data here.

%---------------------------------------------------------------------------------------
% Section Results
%---------------------------------------------------------------------------------------
\section{Results}
\subsection{Clean Images}

% Figure sgra_images
\begin{figure*}[p]
\begin{center}
\begin{tabular}{ccc}
\includegraphics[width=0.3\hsize]{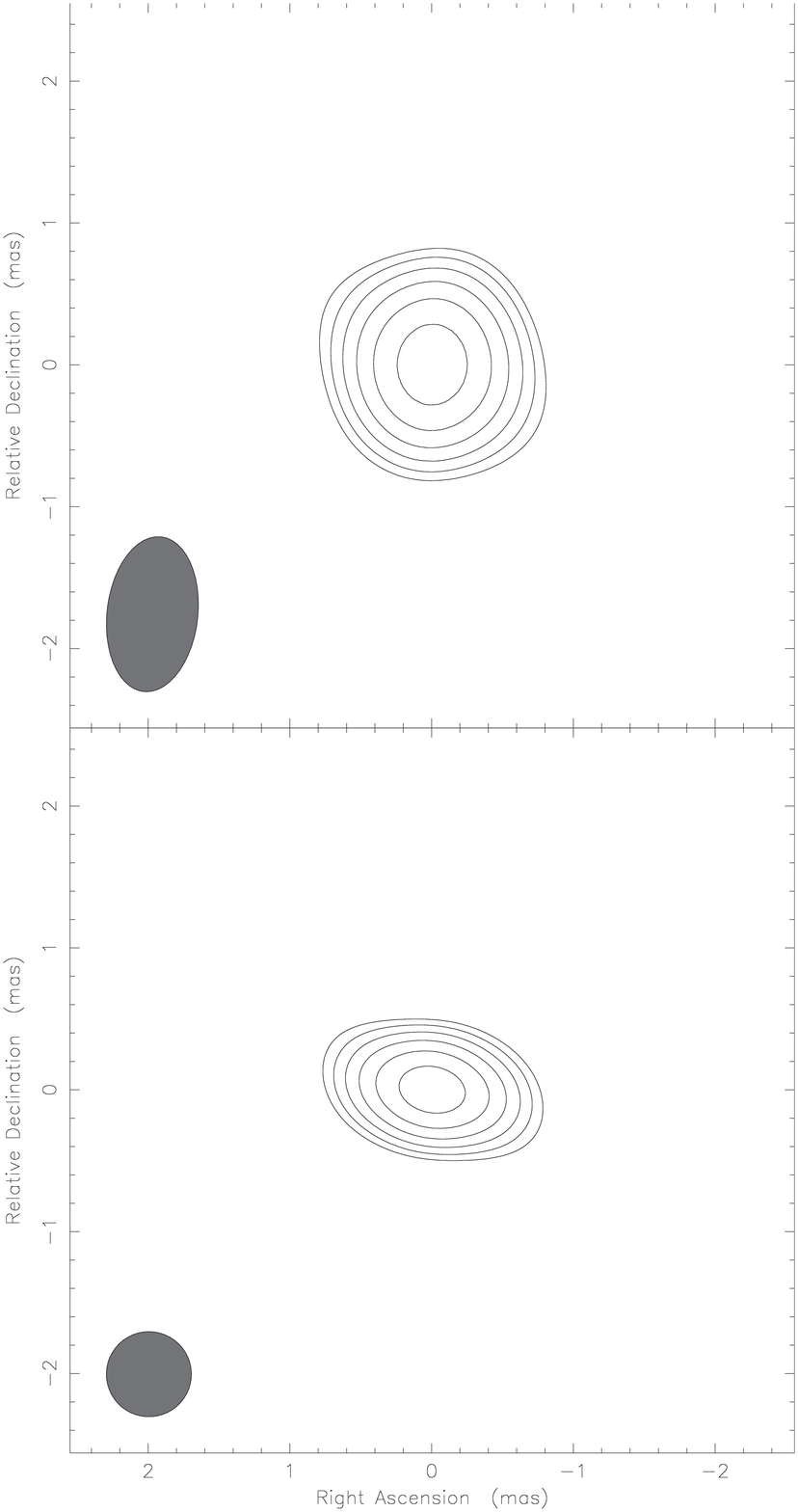}&
\includegraphics[width=0.3\hsize]{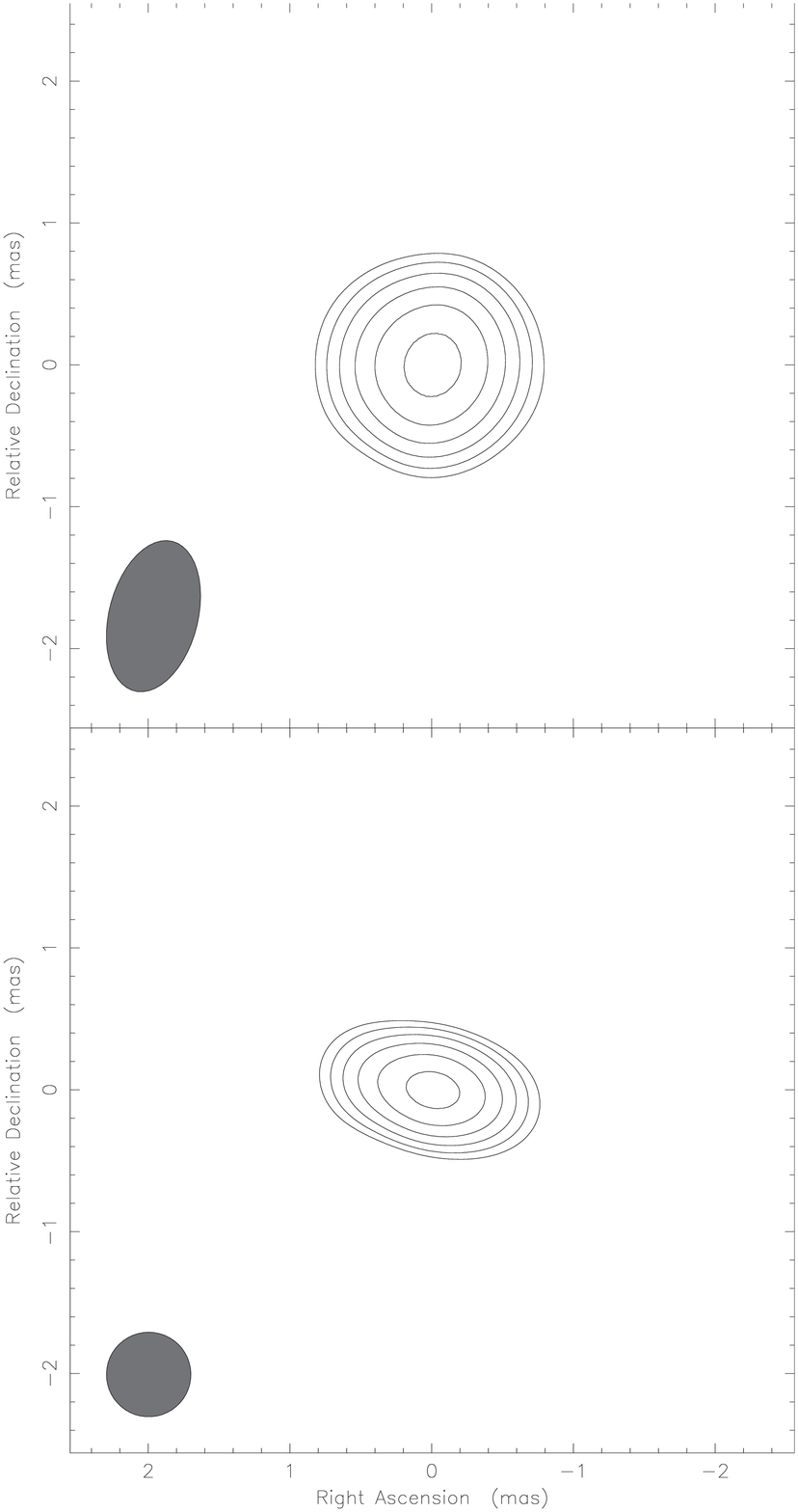}&
\includegraphics[width=0.3\hsize]{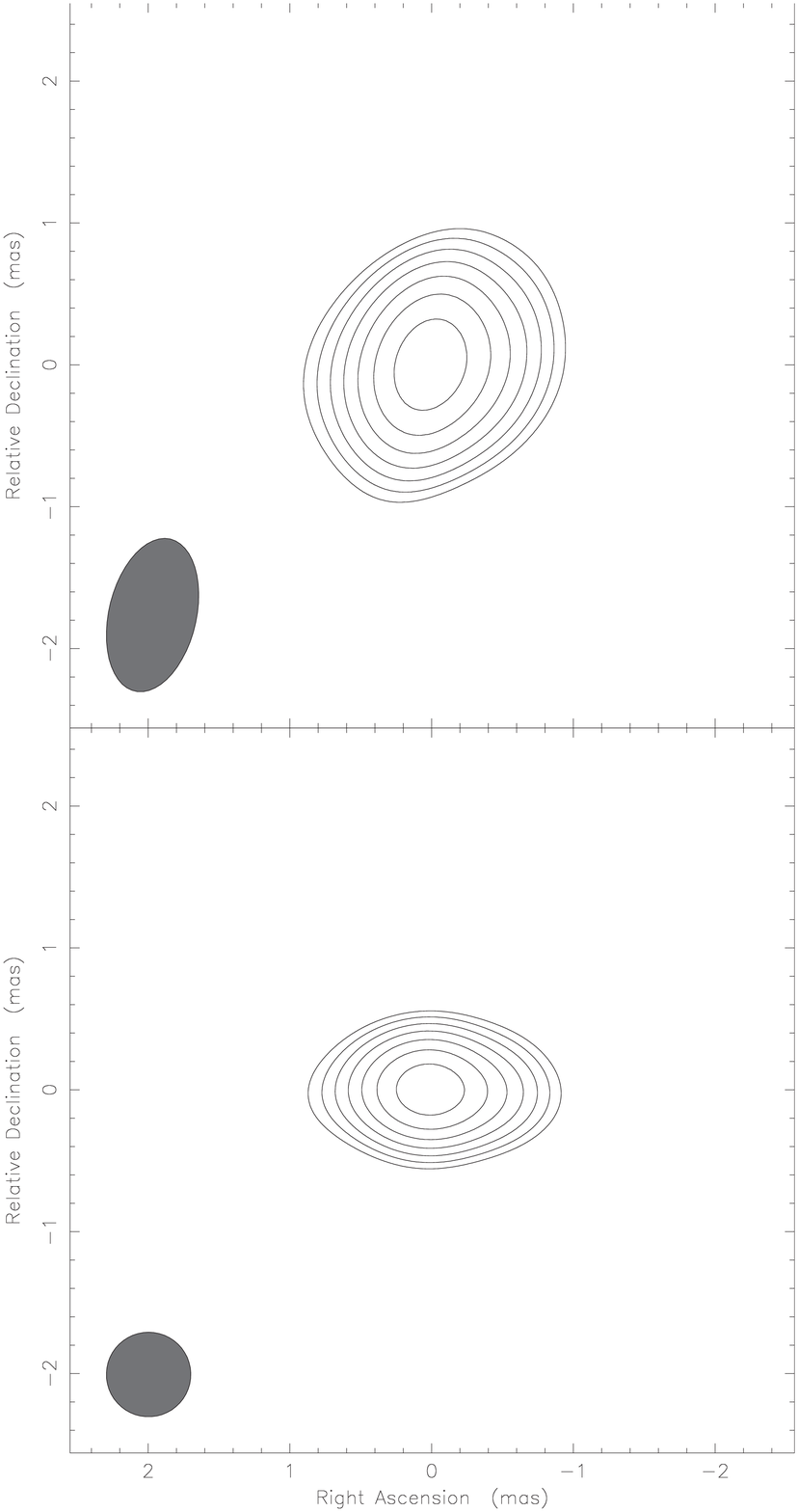}\\
(a)&(b)&(c)
\end{tabular}\\
\vspace{2em}
\begin{tabular}{ccc}
\includegraphics[width=0.3\hsize]{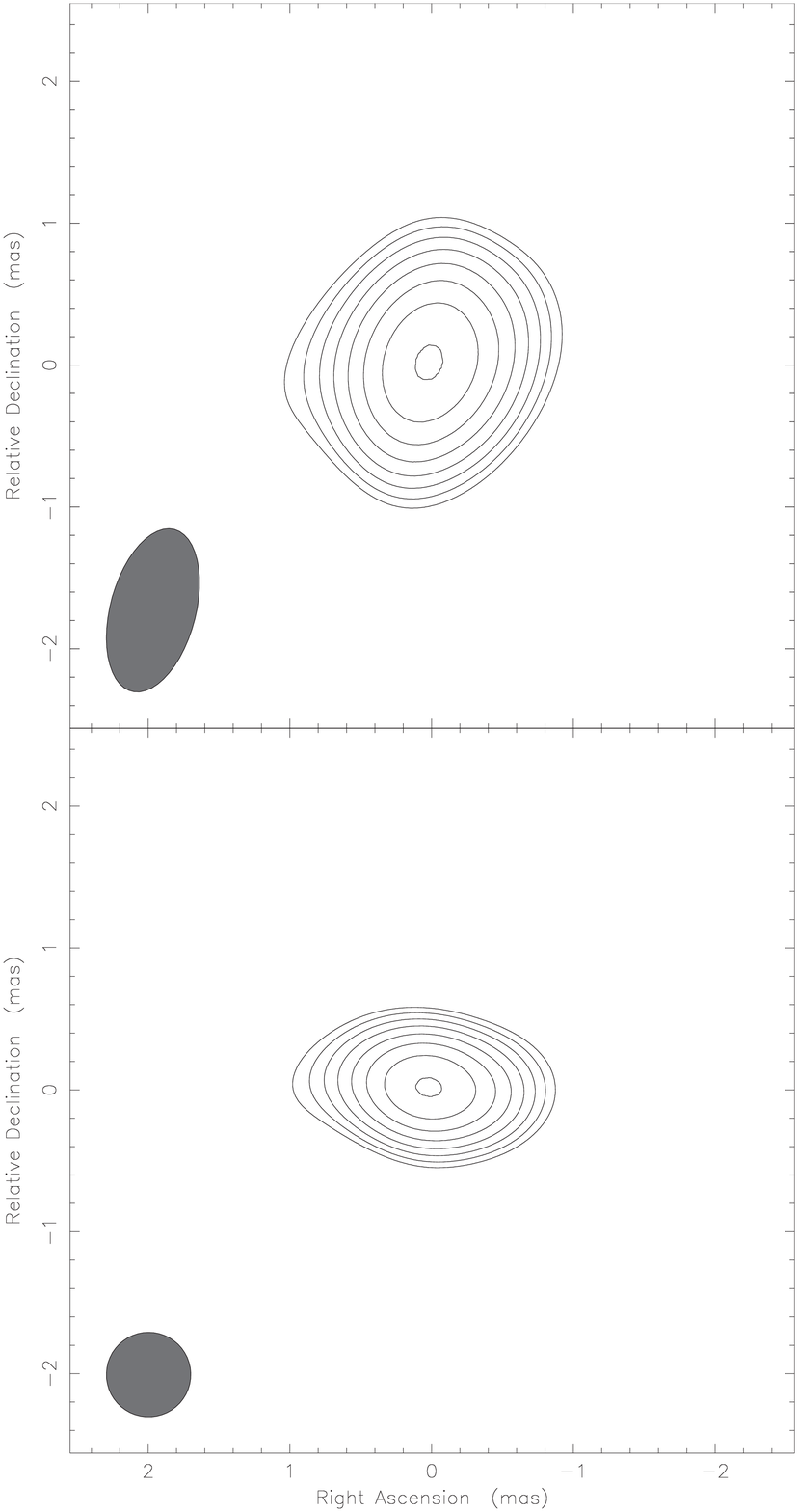}&
\includegraphics[width=0.3\hsize]{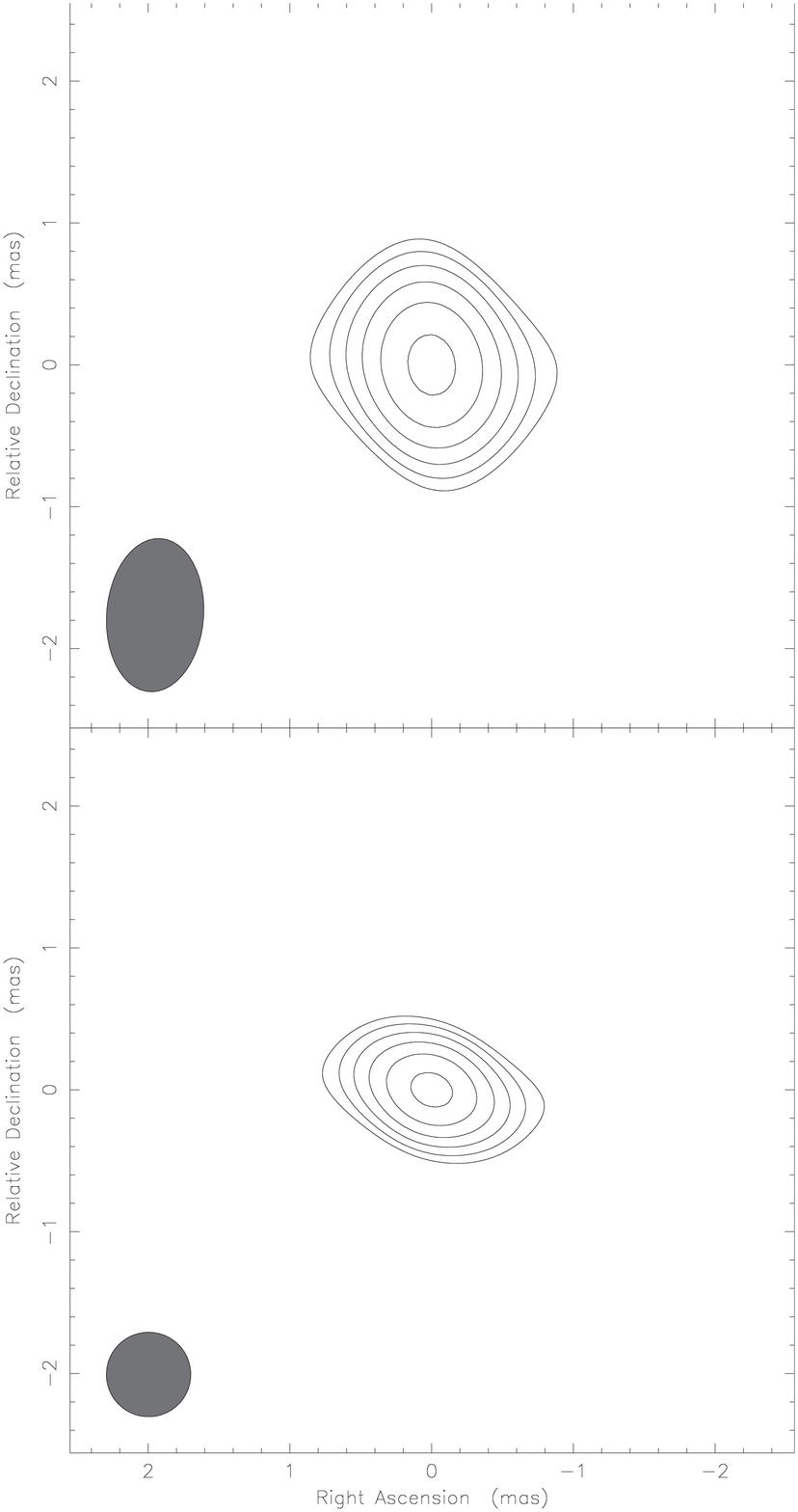}&
\includegraphics[width=0.3\hsize]{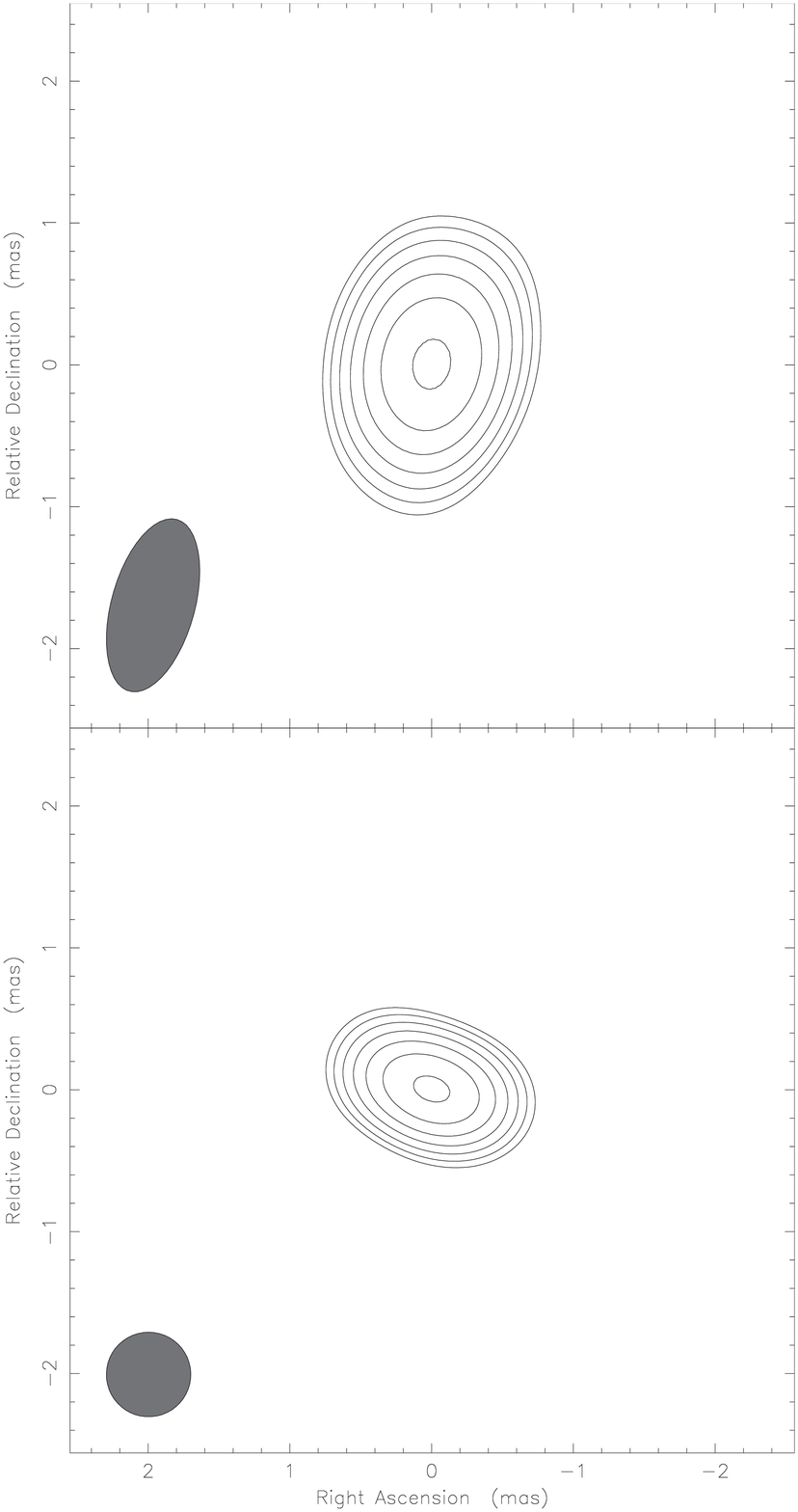}\\
(d)&(e)&(f)
\end{tabular}
\caption{The uniform weighted clean images of Sgr A* (top panels). (a)-(j) are names of epochs shown in Table \ref{tab:CLEAN images}.Bottom panels are super-resolution images a restored circular beam with size of 0.6 mas. The contours are plotted at the level of $5S_{\rm rms}\times\sqrt{2}^n$ ($n=1,2,3,\dots$).
\label{fig:sgra_images}}
\end{center}
\end{figure*}

\begin{figure*}[p]
\begin{center}
\begin{tabular}{cc}
\includegraphics[width=0.3\hsize]{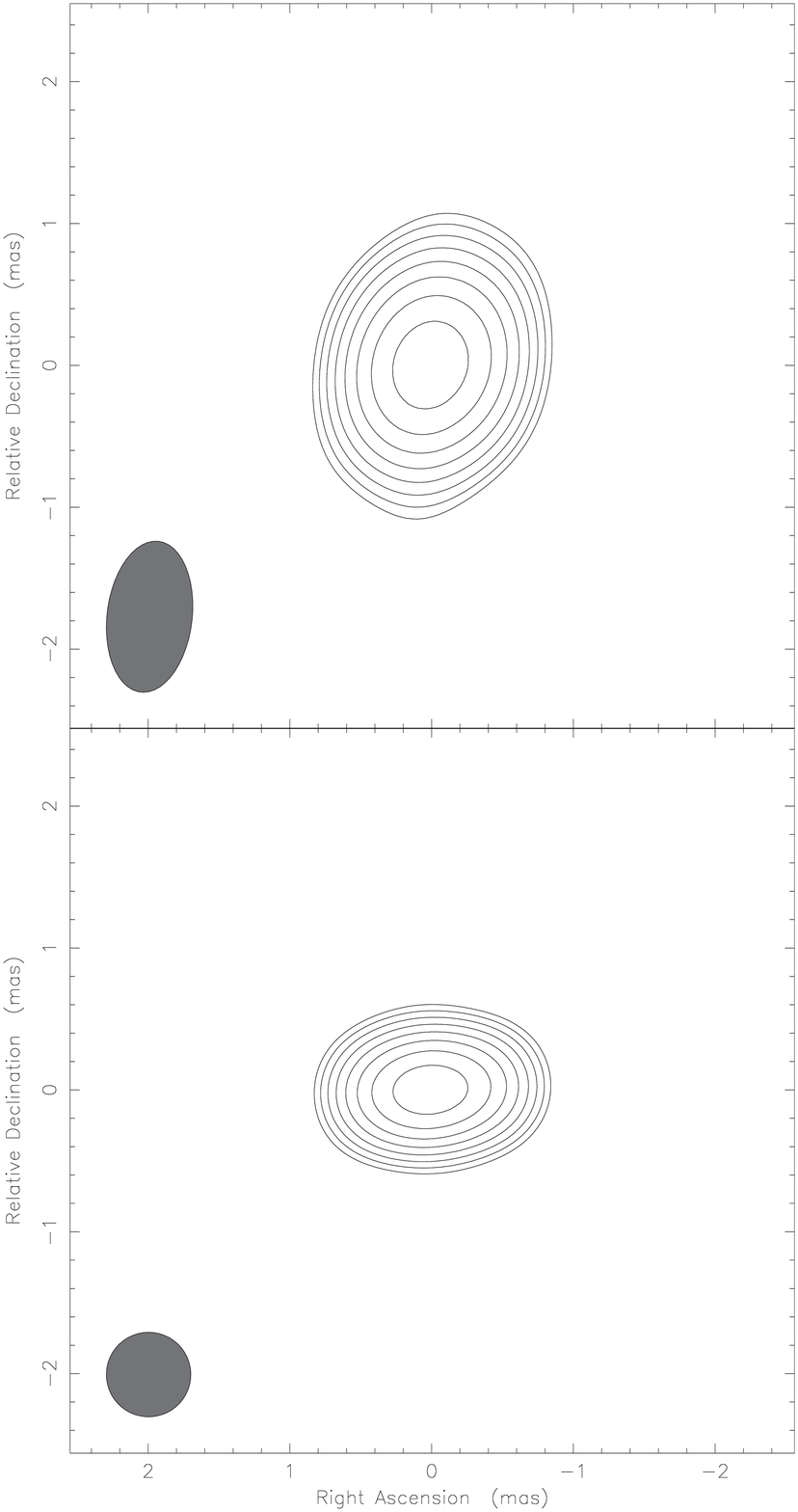}&
\includegraphics[width=0.3\hsize]{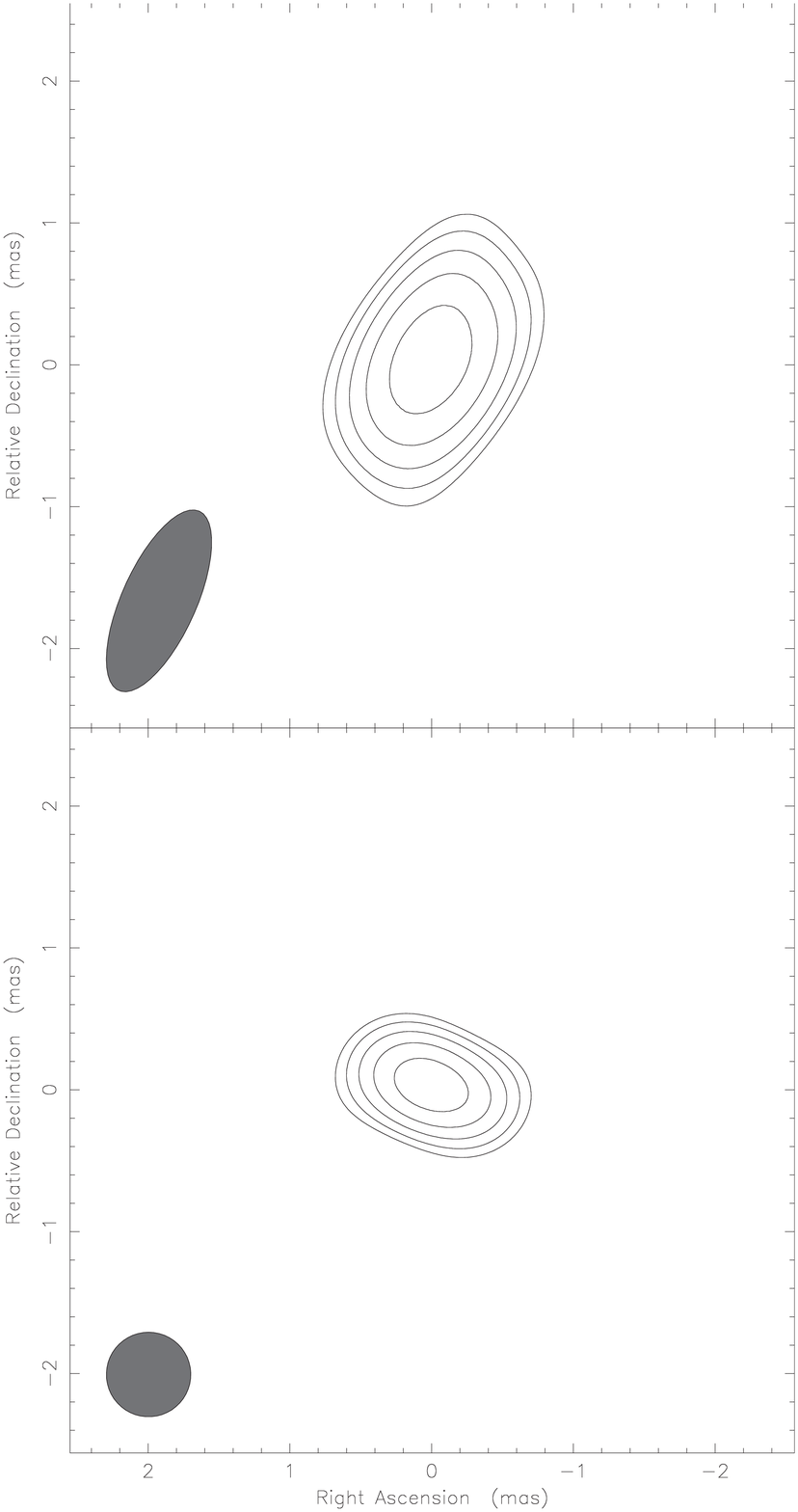}\\
(g)&(h)
\end{tabular}\\
\vspace{2em}
\begin{tabular}{ccc}
\includegraphics[width=0.3\hsize]{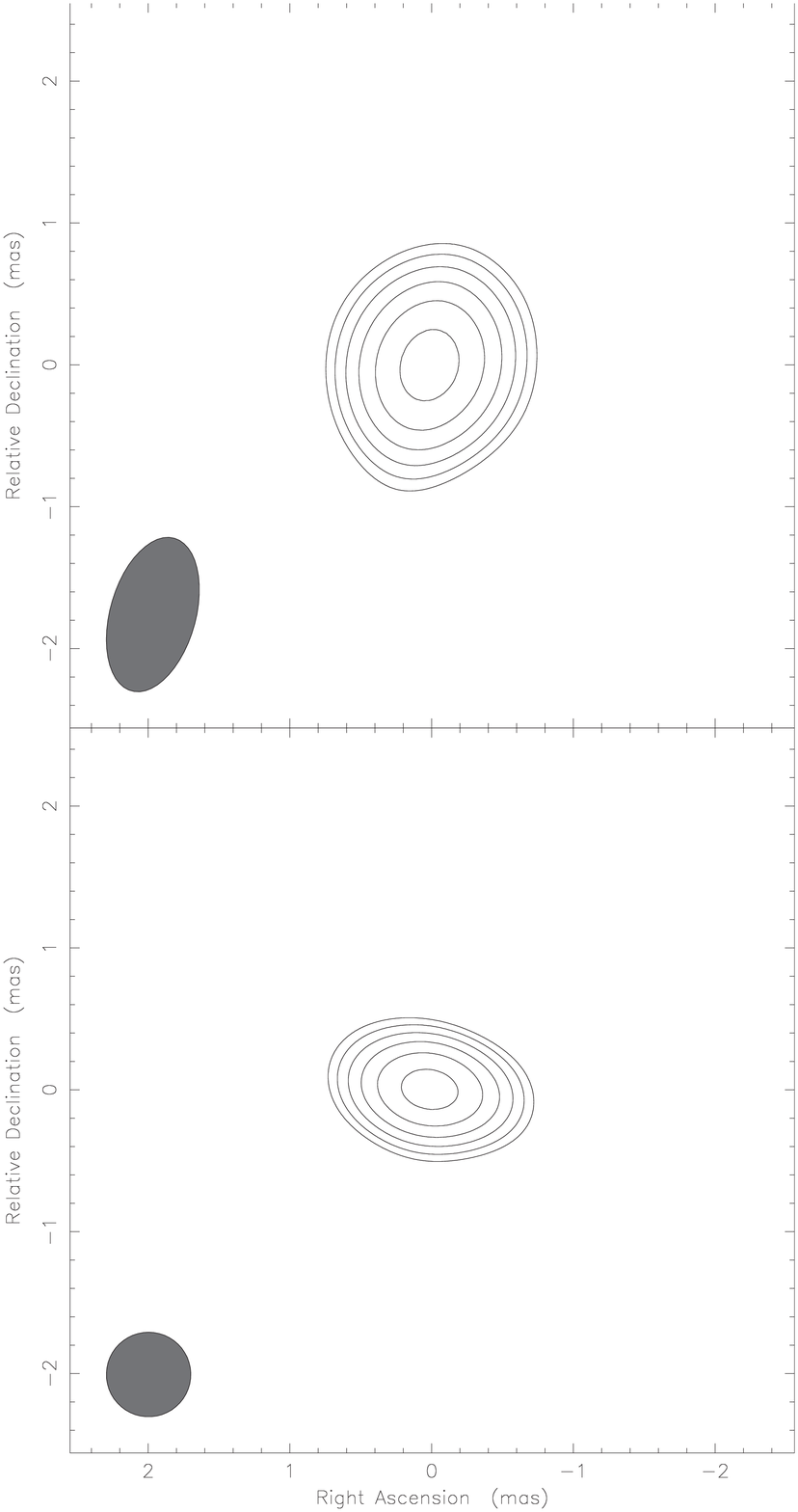}&
\includegraphics[width=0.3\hsize]{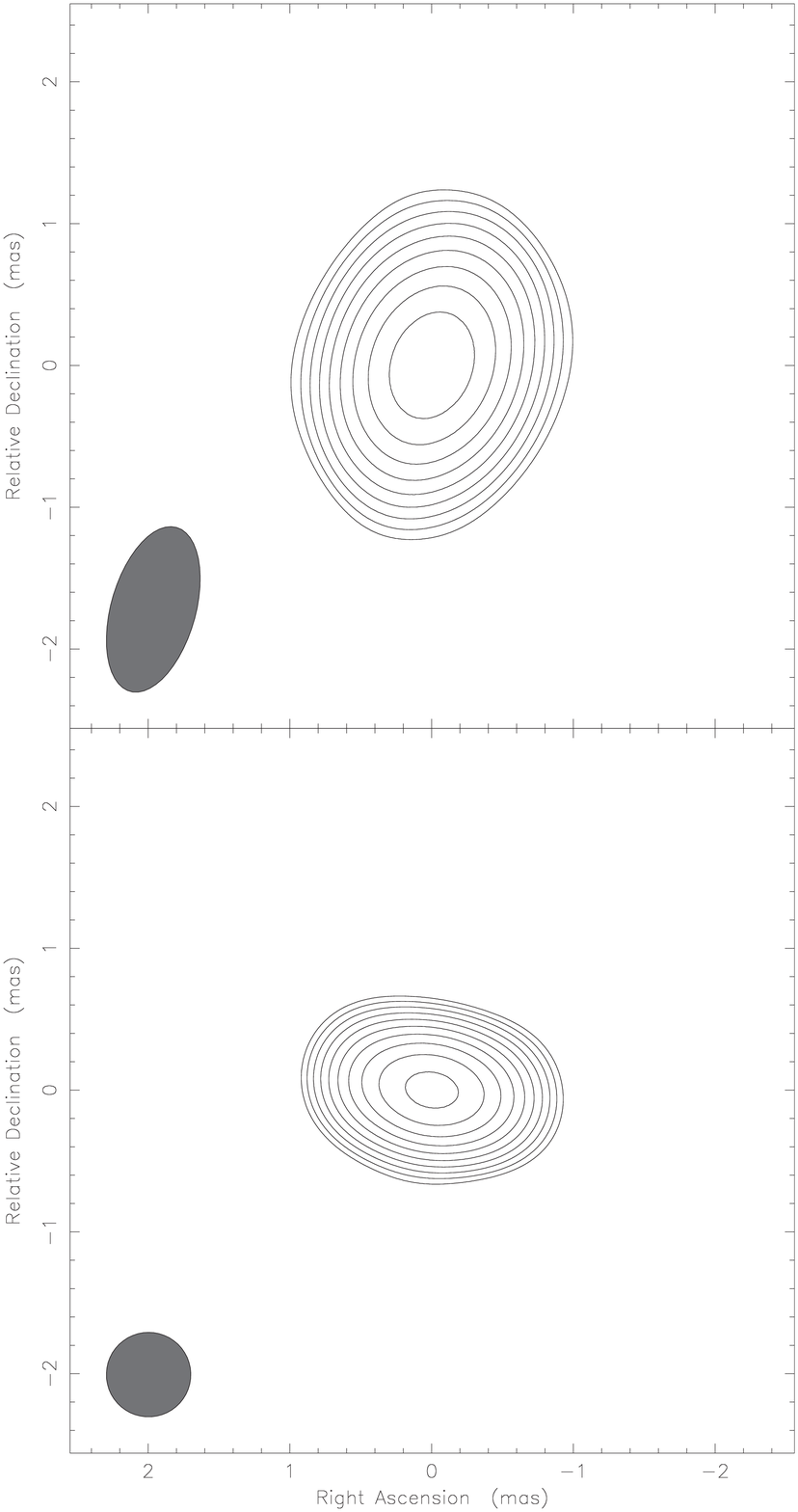}\\
(i)&(j)
\end{tabular}
\addtocounter{figure}{-1}
\caption{Continued.}
\end{center}
\end{figure*}

Figure 5 shows the clean images for all the 10 epochs. The dynamic ranges of the images are more than 30 and the averaged value is about 50. 
As shown in Figure \ref{fig:sgra_images}, Sgr A* has only one component which has nearly symmetric structure. One can also see the symmetry of Sgr A* structure from nearly zero closure phase in Figure \ref{fig:sgra_cpplot_r08310a}. 
One of the most remarkable results in Figure \ref{fig:sgra_images} is that this feature has not changed for about 3 years. In other words, we did not detect ejection of knot-like components through all epochs (although Sgr A* certainly exhibits flux variation during the observed period). 

% Table CLEAN images
\begin{table*}[t]
\begin{center}
\caption{The basic properties of CLEAN images.\label{tab:CLEAN images}}
\begin{tabular}{cccccccccc}\hline \hline
Epoch&Epoch&Epoch&Freq.&$\theta_{\rm maj}$&$\theta_{\rm min}$&PA&$S_{\rm  clean}$&$I_{\rm peak}$&$\sigma_I$\\
(Name)&(DOY)&(yyyy-mm-dd)&(GHz)&($\rm \mu$as)&($\rm \mu$as)&($^\circ$)&(Jy)&(Jy beam$^{-1}$)&(Jy beam$^{-1}$)\\\hline
(a)&2005-294&2005-10-21&43.061&1098&642&6.51&0.85&0.64&0.02\\ 
(b)&2006-079&2006-03-20&43.061&1013&619&4.82&0.63&0.46&0.01\\ 
(c)&2006-109&2006-04-19&43.061&1078&609&-5.18&0.63&0.44&0.01\\ 
(d)&2006-308&2006-11-04&43.061&1126&581&-11.13&0.90&0.62&0.01\\ 
(e)&2007-073&2007-03-14&43.061&1079&683&1.69&0.78&0.59&0.02\\ 
(f)&2007-093&2007-04-03&43.061&1277&568&-15.91&0.93&0.64&0.02\\ 
(g)&2007-264&2007-09-21&43.061&1168&591&-12.39&0.75&0.52&0.01\\ 
(h)&2008-076&2008-03-16&43.061&1327&540&-21.67&0.60&0.42&0.01\\ 
(i)&2008-085&2008-03-25&43.061&1087&580&-14.46&0.62&0.42&0.01\\ 
(j)&2008-310&2008-11-08&43.061&1122&608&-14.15&0.95&0.66&0.01\\\hline
\end{tabular}
\end{center}
%\footnotetext{a}{Sizes and position angles of uniform weighted synthesized beams.}
\end{table*}

We show in Table \ref{tab:CLEAN images} the basic properties of these images, such as the size and position angle of the synthesized beam, the cleaned flux $S_{\rm clean}$, the peak brightness $I_{\rm peak}$ and the image r.m.s $\sigma _{\rm I}$.
One can obviously see the existence of time variations of flux $S_{\rm clean}$.
On the other hand, non-detection of any knot-like structure suggests that the flux variation is associated with the flux change of the single component. 

In previous studies, super-resolution images with restored circular beam are frequently used for examining the resolved structure.
We also show the super-resolution images in the lower panels of Figure \ref{fig:sgra_images}.
One can see that the observed structure of Sgr A* is Gaussian-like rather than point-like.
The Gaussian-like structure of Sgr A* is well-explained by the effect of the interstellar scattering \citep{Narayan1989a,Narayan1989b}.
Sgr A* is well resolved along the east-west direction, since the size of scattering disk along the east-west direction is larger than the north-south direction \citep{Bower2004}.
The size of the lowest contour in super-resolution map is about 1.8 mas in the direction of Right Ascension and about 0.6 mas in the direction of Declination.

These features are consistent with results of previous VLBA observations \citep{Bower1998,Lo1998,Lu2011}.
We note that the number of baselines of VERA is less than VLBA, but the size of the synthesized beam is almost the same as that of previous VLBA observations since long baseline data of VLBA observations are also flagged out because of low S/N ratio.

\subsection{Model-fitting Results}
% Table UVFIT results
\begin{table*}[t]
\caption{The source structure obtained from Gaussian-fitting.\label{tab:UVFIT results}}
\begin{center}
\begin{tabular}{ccccccc}\hline \hline
Epoch&$S_{\rm total}$&$\phi_{\rm maj}$\footnotemark[a]&$\phi_{\rm min}$\footnotemark[a]&PA\footnotemark[a]&$\phi_{\rm int}$\footnotemark[b]&$R_{\rm int}$\footnotemark[b]\\ 
(Name)&(Jy)&($\rm \mu$as)&($\rm \mu$as)&($^\circ$)&($\rm \mu$as)&(R$_{\rm g}$)\\\hline
(a)&1.0$\pm$0.1&740$\pm$10&\verb|<|549&78$^{+1}_{-1}$&340$\pm$20&34$\pm$2\\
(b)&0.8$\pm$0.1&760$\pm$10&\verb|<|507&77$^{+1}_{-1}$&380$\pm$20&38$\pm$2\\
(c)&0.7$\pm$0.1&710$\pm$20&\verb|<|539&86$^{+2}_{-2}$&260$\pm$50&26$\pm$5\\
(d)&0.9$\pm$0.1&710$\pm$10&\verb|<|563&82$^{+1}_{-1}$&260$\pm$30&26$\pm$3\\
(e)&0.8$\pm$0.1&730$\pm$20&\verb|<|540&69$^{+1}_{-1}$&310$\pm$50&31$\pm$5\\
(f)&1.1$\pm$0.1&700$\pm$10&\verb|<|639&71$^{+1}_{-2}$&240$\pm$30&24$\pm$3\\
(g)&0.9$\pm$0.1&700$\pm$10&430$^{+50}_{-70}$&94$^{+4}_{-3}$&240$\pm$30&24$\pm$3\\
(h)&0.8$\pm$0.1&710$\pm$10&350$^{+40}_{-70}$&67$^{+2}_{-2}$&260$\pm$30&26$\pm$3\\
(i)&0.7$\pm$0.1&690$\pm$10&\verb|<|544&79$^{+1}_{-1}$&200$\pm$30&20$\pm$3\\
(j)&1.1$\pm$0.1&700$\pm$10&340$^{+30}_{-40}$&77$^{+1}_{-1}$&240$\pm$30&24$\pm$3\\\hline
$\bar X$ \footnotemark[d]&0.9&720&370&78&270&27\\
$\sigma _X$\footnotemark[e]&0.1&20&40&8&50&5\\\hline \hline
\end{tabular}
\end{center}
\hspace{7em}\footnotemark[a]Sizes and position angles obtained by elliptical-Gaussian-fitting to calibrated visibilities\\
\hspace{7em}\footnotemark[b]Intrinsic sizes of Sgr A*\\
\hspace{7em}\footnotemark[c]time-averages of each parameter\\
\hspace{7em}\footnotemark[d]Standard deviations of each parameter\\
\end{table*}

\begin{figure*}[!th]
\begin{center}
\begin{tabular}{cc}
\includegraphics[width=0.4\hsize]{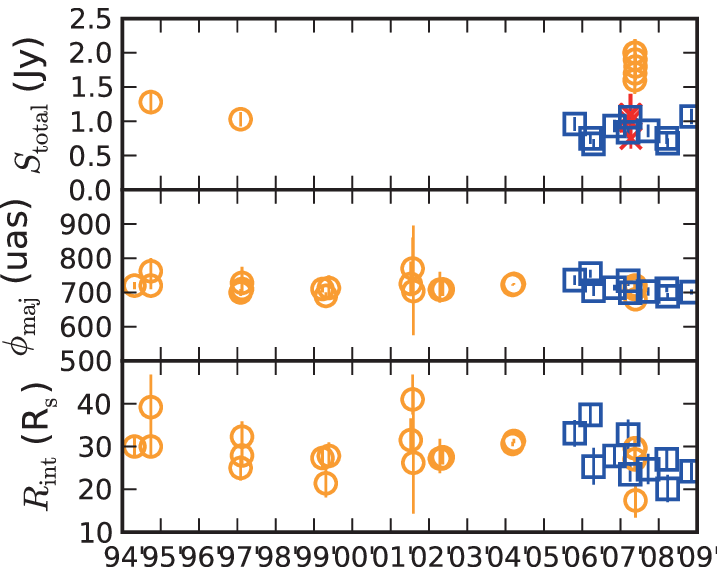} & \includegraphics[width=0.4\hsize]{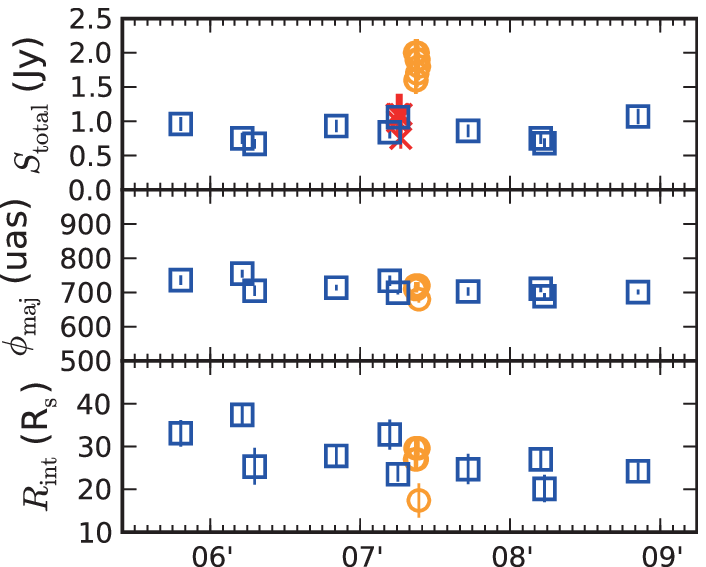}\\
(a) & (b)
\end{tabular}
\caption{The time variations of the flux and structure of Sgr A*. From the top of figure, vertical axes are correspond to the flux, the size of Major axis and the intrinsic size. The squares (colored blue in the on-line version) are our results, while the circles (colored orange in the on-line version) and the crosses (colored red in the on-line version) are VLBA results. The circled mark are the result of \citet{Yusef2009} and the error bar indicates the range of the intra-day variability. (a) shows all data shown in Table \ref {tab:UVFIT results} and \ref{tab:VLBA results}, while (b) shows VERA data and VLBA data observed around VERA epochs (2005-2008).
\label{fig:time variation}}
\end{center}
\end{figure*}

In previous studies of Sgr A*, VLBI-scale source structure has been quantified by using Gaussian fitting to visibilities, since the observed Sgr A* structure is well explained by an elliptical Gaussian distribution \citep{Bower2004,Shen2005,Lu2011}.
In fact, our results also show a symmetric and Gaussian-like structure consisting of a single component, as described in the previous subsection.
Hence, we fitted Gaussian models directly to the calibrated visibility data with the least-square method as described in Section 2. Figure \ref{fig:time variation} shows their time variations. 
The parameters obtained from the fitting, such as total flux $S_{\rm total}$, size of the major/minor axes $\phi_{\rm maj}$/$\phi_{\rm min}$, position angle $PA$ and intrinsic size $\phi _{\rm int}/R _{\rm int}$ are summarized in Table \ref{tab:UVFIT results}. 
In addition to the fitting results of individual parameters, we show time-averages and standard deviations of each parameter throughout all the epochs at the bottom line of Table \ref{tab:UVFIT results}.
As an example of Gaussian fitting, we have already shown the results for epoch (g) in Figure \ref{fig:sgra_projplot_r08310a}.

We note about the upper limits of the minor axis size of Sgr A* set in several epochs and the errors on PAs in Table \ref{tab:UVFIT results}.
In some epochs, upper limits of the minor axis size are obtained.
In these epochs, we could not detect structure elongated in the direction of the minor axis of Sgr A*.
The most probable reason is the lack of spatial resolution; the major axis of the synthesized beam beam is oriented in a N-S direction closer to the direction of the minor axis of Sgr A* and major size of the beam is typically $\sim $ 1.2 mas, which is $\sim$ 3 times greater than typical minor axis size of Sgr A*.
In addition to the lack of spatial resolution, the poor UV coverage in the N-S direction probably makes the measurement of minor axis size of Sgr A* more difficult.
Therefore, in these epochs, we set a half of the major axis size of the synthesized beam ($\sim$ 0.6 mas) as an upper limit of the minor axis size of Sgr A*. This upper limit (typically 0.5-0.6 mas) is reasonable, because the Gaussian distribution with the minor axis size of 0.5-0.6 mas should be resolved-out in our observations. We confirmed that if the minor axis size was 0.6 mas, fringes would have been detected only in Mizusawa-Ogasawara and Iriki-Ishigaki baselines considering the sensitivity of our observations, which is clearly inconsistent with our results. This fact strongly indicates the minor axis size was smaller than 0.6 mas.

% Figure time variation

%We also added the results of \citet{Yusef2009} {\bf in Figure \ref{fig:time variation} in addition to data in Table \ref{tab:VLBA results}}. \citet{Yusef2009} estimated the intra-day variability of the Sgr A*'s flux by fitting the Gaussian model with the fixed sizes derived in \citet{Bower2004}. Therefore, in Figure \ref{fig:time variation}, we show only the ranges of the flux variations within each day, which is the only available information in \citet{Yusef2009}.

We successfully measured the major-axis size $\phi _{\rm maj}$ through all the epochs.
The time-average of $\phi _{\rm maj}$ is 720 \uas, which is consistent with previous VLBA observations \citep{Bower2004,Lu2011}.
On the other hand, we could not determine the minor-axis size $\phi _{\rm min}$ in several epochs.
The time-average of $\phi_{min}$ and PA are about 370 \uas and 78$^\circ$ (without using upper limit data), and also similar to previous results of VLBA.
However, the minor axis size and also position angle of Sgr A* are possibly affected by synthesized beam (i.e. uv coverage and spatial resolution).
Thus, it is difficult to discuss the time variation of the minor axis size and position angle.
Hence, in this paper, we concentrate on the variation of the flux and major axis size.

% Figure flux vs size
%\begin{figure}
%\begin{center}
%\includegraphics[width=1.0\hsize]{figure7.eps}
%\caption{The relation of the total flux and the intrinsic size of Sgr A*. The squares (colored blue in the on-line version) indicate our results, while the circles (colored orange in the on-line version) indicate VLBA results. The parabolic dashed-lines are the contour of brightness temperature calculated assuming $\phi_{\rm int\_ maj}=\phi_{\rm int\_ min}=\phi _{\rm int}$ in (\ref{eq:Tb}). The contours are plotted at the level of $5n\,\times\,10^9$ K (for n=1,2,\dots,13).
%\label{fig:flux vs size}}
%\end{center}
%\end{figure}

Remarkably, in addition to the flux variation, one can see the variation of the intrinsic size.
The standard deviation of $S_\nu$, 0.1 Jy corresponds to $\sim$ 11 \% of the time-averaged flux $\bar S_\nu$, while that of $\phi _{\rm int}$, 50 \uas corresponds to $\sim$ 19 \% of the time-averaged intrinsic size $\bar \phi_{\rm int}$. We confirmed the significance of time variations in $S_{\rm total}$, $\phi_{\rm maj}$ and $\phi_{\rm int}$ from $\chi ^2$ tests of hypotheses of no time variation in those quantities. The error weighted averages of $S_{\rm total}$, $\phi_{\rm maj}$ and $\phi_{\rm int}$ are 0.9 Jy, 710 \uas and 290 {\uas}, resulting $\chi ^2$s based on the hypotheses are 20, 41 and 46 respectively. The $\chi ^2$ tests based on those $\chi ^2$s indicate that the hypotheses are rejected at the significance level of 5 $\%$.

%However, To statistically confirm the brightness temperature variation, we carried out the chi-square-test of the brightness temperature, and confirmed that the hypothesis of a constant brightness temperature can be rejected at a significance level of 99 $\%$.

A noteworthy feature in Figure \ref{fig:time variation} is that the total flux of Sgr A* was flared up to about 2 Jy in May 2007 \citep{Lu2011}. They also reported that the fluxes at 22 GHz and 86 GHz are high in these epochs as well. 
Compared with the results of a dense monitor in \citet{Herrnstein2004}, observed fluxes of Sgr A* are relatively high.
In the present paper, we refer to this flaring of Sgr A* as {''}the flaring event{''}.

Combination of our and VLBA results suggests that the duration of the flaring event was longer than 10 days and shorter than 151 days. The time-averaged flux during the flaring event is 1.79 $\pm$ 0.05 Jy. This is about 1.7 times higher than epoch (f) ($\sim$ 1 month prior to the flaring event), and also 2.1 times higher than epoch (g) ($\sim$ 6 months posterior to the flaring event). Meanwhile, the time-averaged intrinsic size during the flaring event is 267 $^{+27}_{-27}$ $\rm \mu$as, which is consistent with those of epochs (f), (g) and the time-averaged value of our results within 1 $\sigma$ level. This result indicates that the brightness temperature of Sgr A* increased during the flaring event (see \S4.1).

\section{Discussion}
%First, we consider a case of a circular shape whose radius corresponds to the intrinsic major axis size of Sgr A* (i.e. $\phi_{\rm int\_ maj}=\phi_{\rm int\_ min}=\phi _{\rm int}$). This would not be an unrealistic estimate because the intrinsic size of the minor axis is reported as being about 260 \uas and comparable with that of the major axis \citep{Bower2004}. Here, we call the brightness temperature calculated under this assumption as $T_{b1}$ and shows it in the upper panel of Figure \ref{fig:Tbxphi_int_min}. One can see that the scatter of observed flux and intrinsic size requires a change in $T_{b1}$ ranging from $\sim$ 5 $\times$ 10$^9$ K to $\sim$ 19 $\times$ 10$^9$ K. $\chi ^2$ test for the brightness temperature shows that a hypothesis of a constant temperature was rejected at the significance level of 99 $\%$, indicating the existence of the time variation in the brightness temperature under this assumption.

% Minor Axis size kotei 
%Second, we consider a case of fixing the minor axis size. We shows the brightness temperture $T_b2$ which was calculated under the assumption of $\phi_{\rm int\_ maj}=\phi _{\rm int}$ and $\phi_{\rm int\_ min}=260 \, {\rm \mu as}$ in the bottom panel of Figure \ref{fig:Tbxphi_int_min}. $T_b2$ ranges from 7 $\times$ 10$^9$ K to 17 $\times$ 10$^9$ K. We also carried out the $\chi ^2$ test, and confirmed that a hypothesis of a constant temperature was rejected at the significance level of 99 $\%$. Thus, the time variation in the brightness temperature is indicated also under this assumption.
\subsection{Variability of the brightness temperature}
% Figure time variation
\begin{figure}[!t]
\center
\includegraphics[width=0.8\hsize]{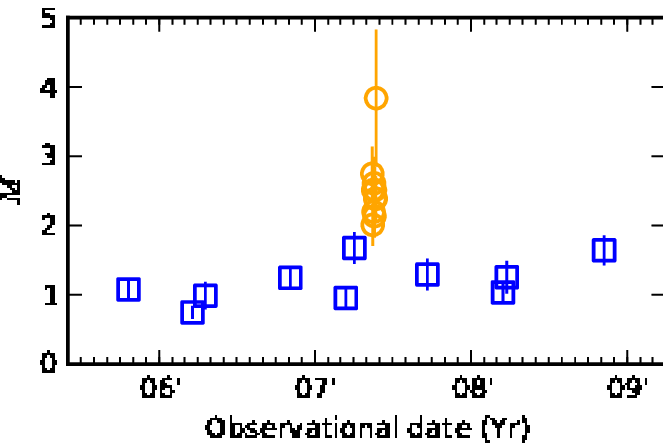}
\caption{The time variations of the multiply of the intrinsic-minor-axis size and the brightness temperature.
\label{fig:Tbxphi_int_min}}
\end{figure}

The total flux and the major axis size of Sgr A* were successfully measured and then they can be used for discussing the possibility of the variation in the brightness temperature $T_b$, which is calculated by
\begin{eqnarray}
T_b &=& \frac{2c^2}{k_B \nu ^2} \frac{S_{\rm total}}{\pi \phi_{\rm int\_ maj} \phi_{\rm int\_ min}}\\\label{eq:Tb}
    &=& 9.54 \times 10^{8} \,{\rm K}\times \left( \frac{\nu}{\rm 43 \,GHz} \right)^{-2}\nonumber\\
    &&\left( \frac{S_{\rm total}}{\rm 1\, Jy}\right)\left( \frac{\phi_{\rm maj\_ int}}{\rm 1 \, mas}\right)^{-1} \left( \frac{\phi_{\rm min\_ int}}{\rm 1 \, mas}\right)^{-1}.
\end{eqnarray}
$\phi_{\rm int\_ maj}$ and $\phi_{\rm int\_ min}$ are intrinsic sizes of the major and minor axes. Here, the brightness distribution is set to be an uniform elliptical disk. Its diameter is set to be FWHM sizes of the elliptical Gaussian model, ant its total flux equals to that of the elliptical Gaussian model.

% Minor Axis Free
Since the minor-axis size was not determined in many epochs, we define the multiple of the brightness temperature and the minor axis intrinsic size given by 
\begin{eqnarray}
M&\equiv &\left( \frac{T_b}{10^9 \rm K} \right) \left( \frac{\phi_{\rm min\_int}}{260\, {\rm \mu as}} \right)\\
&=& 3.67 \left( \frac{\nu}{\rm 43 \,GHz} \right)^{-2}\left( \frac{S_{\rm total}}{\rm 1\, Jy}\right)\left( \frac{\phi_{\rm maj\_ int}}{\rm 1 \, mas}\right)^{-1},
\end{eqnarray}
where the normalization factor of 260 ${\rm \mu as}$ for the intrinsic-minor-axis size is referred to from averaged value of VLBA measurements in \citet{Bower2004}. We show the time variation in $M$ in Figure 7. If the brightness temperature does not varies with time, the variation in $M$ should originate from that in the intrinsic minor size. Here, we assume the time-averaged value of the intrinsic-minor-axis size for our VERA data to be 260 \uas.

In our VERA results (without VLBA data), the constant brightness temperature requires variation in observed-minor-axis size ranged from $\sim 190$ \uas to $\sim 415$ \uas. It would be difficult to detect such a variation with VERA considering the beam size along the N-S direction. Thus, the variation in $M$ could be explained only by the variation in the intrinsic-minor-axis size.

On the other hand, the combination of VERA data and VLBA data requires the increase in the brightness temperature at the flaring event. The error-weighted averages of $M$ are 1.0 for our VERA data and 2.3 for VLBA data at the flaring event. If there was no time variation in the brightness temperature through our VERA data and the flaring event, the observed-minor-axis size of $590$ \uas should have been required at the flaring event. However, \citet{Lu2011} reported the time-averaged minor axis size of 400 $\pm$ 10 \uas at the flaring event and inconsistent with predicted observed-minor-axis size of $590$ \uas. Thus, the increase in $M$ at the flaring event can not be explained only by variation of the intrinsic-minor-axis size, and requires the increase in the brightness temperature.

\subsection{The origin of variability of the brightness temperature}
\subsubsection{Possibilities of an extrinsic origin}
The variation of the total flux and intrinsic size of Sgr A* requires most likely requires the variation of the brightness
temperature. Before discussing this, here, we comment on the possibility of interstellar scattering as an origin of the time variations of Sgr A*. Since the detailed structure of the scattering medium is still unknown, it is not possible to totally rule out the effect of interstellar scattering. The scattered image of Sgr A* is thought to be created by turbulent plasma along the line of sight. A typical time scale for the scattering is given by refractive time scale, which is determined by relative motions of the observer, turbulent plasma and background source. The refractive time scale for Sgr A* is estimated to be $\sim$ 0.5 $\lambda ^2$ year cm$^{-2}$ , given a relative velocity of 100 km s$^{-1}$ \citep{Narayan1989a,Bower2004}. At 43 GHz, the refractive time scale becomes $\sim$ 3 months. Most of our observations are separated by periods longer than 3 months, implying that the time variations of the total flux may be partly caused by refractive changes. However, epochs (e), (f) and the flaring event reported by \citet{Lu2011} are separated by only a month (see Figure \ref{fig:time variation}). Furthermore, \citet{Lu2011} reported that the spectral index became harder than previous results. In simple models of interstellar scattering, the modulation index of the flux density decreases with frequency in a strong scattering regime \citep{Rickett1990}. This would lead to an anti-correlation between spectral index and mm-flux. Therefore, the flaring event is inconsistent with the effects of simple models for interstellar scintillation. Thus, observed variation of brightness temperature is likely intrinsic. In the following subsections, assuming that the flaring event in \citet{Lu2011} is intrinsic, we discuss possible intrinsic origins of the flaring event in 2007.  

\subsubsection{Models with Jets, Expanding Plasmon or Hot Spot}
Here, we discuss whether the brightness temperature variation associated with the flaring event can be explained by models suggested for intra-day variations. In previous studies, several models were proposed as the origin of intra-day variation, such as models with jet \citep{Falcke2009,Maitra2009}, expanding plasmon model \citep{Yusef2008,Yusef2009} and orbiting hot-spot model \citep{Broderick2006,Eckart2006,Eckart2008}.

Both VERA and VLBA have not detected any additional components in the Sgr A* images around the flaring event in 2007. From this fact, an ejection of a sub-relativistic or relativistic bright component \citep{Falcke2009,Maitra2009} would be ruled out in this case. The size of Sgr A* at 43 GHz is about 0.7 mas along the major axis, corresponding to a light crossing time of $\sim$ 46 min. If the flaring event is owing to an ejection of an outflow or a jet, the velocity of the new component must be less than 0.003 c so that the new component cannot be resolved from the persistent component within 10 days. This velocity upper-limit is significantly smaller than the predicted value of 0.1c in \citet{Falcke2009}.

The expanding plasmon model \citep{van1966}, in which creation of an expanding plasma blob causes flux variation, predicts an expansion velocity comparable to the upper limit of 0.003 c in case of the intra-day variation of Sgr A* \citep{Yusef2008,Yusef2009,Li2009}. However, when the expansion velocity is comparable to 0.003 c, the adiabatic cooling time-scale is shorter than a day. If the expansion velocity is smaller than 0.003 c, the adiabatic cooling time-scale can be longer. However, in such a case, the synchrotron cooling dominates the adiabatic cooling. According to \citet{Marrone2008}, the synchrotron cooling time-scale is given by
\begin{eqnarray}
t_{\rm syn}=39 \,{\rm hrs}\, \left( \frac{\nu}{86\,{\rm GHz}}\right)^{-1/2} \left(\frac{B}{10 \,{\rm G}} \right)^{-3/2}. \label{eq:syntime}
\end{eqnarray}
In previous studies, typical magnetic-field strength is estimated to be around or larger than 10 G \citep{Yusef2008,Yusef2009,Li2009}. If the magnetic-field strength of 10 G is adopted, the flux decay time-scale at 86 GHz is much shorter than 10 days, which contradicts the results of \citet{Lu2011}.

The observed VLBI structure presumably does not favor the expanding plasmon model. As described in Section 3, the total flux of 22 GHz also increased in the flaring event \citep{Lu2011}. If the expanding plasmon is the origin of the flaring event, this hot plasma should be generated outside of the 22-GHz photosphere. A typical intrinsic size of Sgr A* is 74 R$_{\rm s}$ or 0.73 mas at 22 GHz \citep{Falcke2009}. If there is a bright component outside of the radio-photosphere at 22 GHz, its should be resolved in VLBI maps at 43 GHz or 86 GHz, being inconsistent with the results in the present paper. For the same reason, it is unlikely that the flux increase is caused by an orbiting hot spot \citep{Broderick2006,Meyer2006a,Trippe2007,Eckart2006,Eckart2008, Li2009}.

In summary, the models with new components emerged in optically-thin regions would not be able to explain the flaring event. Finally, we note that the persistent jet model \citep{Falcke2000,Yuan2002,Markoff2007}  is not ruled out. Most likely, it is difficult to explain such a flux variation by newly emerged components, but it could be explained assuming that, for example, the flux variation is owing to the variation of the photosphere of the jet core.

\subsubsection{Models with RIAF disk}
The models with additional bright components would be not suitable for explaining the flaring event. Hence, the flaring event is likely to be associated with a brightness increase of the photosphere. Compared with the light crossing time and Keplerian orbital period, the flaring event had much longer duration. This fact suggests that the flaring event is likely owing to an event that establishes a new steady state in Sgr A*'s accretion disk. The existence of the high density state of Sgr A* is also suggested by \citet{Herrnstein2004} based on the preliminary bimodal distribution of flux density. Moreover, recent Event Horizon Telescope observations \citep{Fish2011} found also such an establishment of new steady state with higher brightness temperature at 1.3 mm. Thus, the flaring event may be possibly caused by such a new steady state.

% Accretion Disk model %
According to accretion disk models, an electron energy distribution and its spatial dependency are crucial for determining the radio emission of Sgr A*. However, the detailed properties of these have been unclear. For instance, some models succeeded in explaining the radio spectrum of Sgr A* by emission from thermal electrons \citep{Kato2009}, while other models fail to explain the radio spectrum of Sgr A* using only thermal electrons \citep{Moscibrodzka2009} and need non-thermal electrons \citep{Yuan2003,Broderick2009,Broderick2010}. In fact, recent theoretical studies imply that true electron distribution functions contain non-thermal components \citep{Riquelme2012}. Thus, at present we should consider both thermal and non-thermal electrons for the radio emission of Sgr A*.

Thermal electrons are probably not suitable for explaining the flaring event based on a self-similar RIAF model.  If the flaring event is established by emission from thermal electrons, an increase in brightness temperature means increase in the electron temperature, since thermal synchrotron emission at 43 GHz is thought to be optically thick. According to studies of
self-similar ADAF model \citep{Mahadevan1997}, an electron temperature $T_e$ in ADAF has a dependency of $\propto \mdot ^{-1/14}$, where $\mdot $ is a mass accretion rate. This dependency is not changed even in the self-similar ADIOS model. Thus, an increase in the electron temperature needs a larger decrease in the mass accretion rate. On the other hand, because the luminosity of a RIAF disk is proportional to $\sim \mdot ^2$ \citep{Kato2008}, the radio flux should decrease as the mass accretion rate decreases. This is inconsistent with our observational results that both the brightness temperature and the total flux increased at the flaring event. Thus, based on theoretical studies of self-similar RIAF, thermal electron origin is not favored. However, since the electron temperature profile, which seriously affects the radio emission, is poorly constrained, our results might be explained by a new steady electron temperature profile, for instance, owing to standing shock in accretion disk \citep{Nagakura2010}. Moreover, the recent numerical GRMHD simulations suggest that a scaling law between the radio flux and the mass accretion rate is slightly different from that of self-similar RIAF model \citep{Moscibrodzka2012}. Future numerical studies might explain this flaring event with only thermal electrons.

% Non-thermal electrons %
Non-thermal electrons could be the origin of that. Recent axis-symmetric particle-in-cell (PIC) simulation of the collision-less magnetic-rotational-instability (MRI) shows that MRI in RIAF causes magnetic re-connections and that non-thermal components can be produced by magnetic re-connections \citep{Riquelme2012}. However, the detailed production mechanism of stationary non-thermal
components has been unclear. At least, a single event of electron acceleration is not favored (\S 4.2.1), but the mechanism of continuous injection of accelerated electrons should be required. A standing shock in the accretion disk \citep{Nagakura2010} might explain continuous acceleration of electrons.

In summary, as the origin of the flared emission at the flaring event, thermal electrons is not thermal electrons are not favored based on a self-similar RIAF model. However, an establishment of a new steady electron temperature profile or future numerical GRMHD simulation might explain the flaring events. Non-thermal electrons could explain the flaring event assuming a continuous production mechanism of non-thermal electrons such as a standing shock in accretion disk. Simultaneous measurements of detailed radio spectra varied from cm to sub-mm wavelengths would be helpful to constrain the properties of electron distribution at the flaring event.

\section{Summary}
We presented the results of multi-epoch VERA observations of Sgr A* at 43 GHz performed from 2005 to 2008. Observed images showed that Sgr A* has only one component for about 3 years and we did not detect any significant structural changes such as a creation of new jet components through all epochs. Based on the analyses with Gaussian fitting to the observed visibilities, we detected time variations of the total flux and the intrinsic size. By combining the results of VLBA and VERA observations, we found the flaring events at least longer than 10 days. Furthermore, we succeed in determining the intrinsic size before/after the flaring event. Our measurements indicate that the intrinsic sizes remained unchanged within 1 $\sigma$ level from the sizes before/after the flaring event, indicating that the brightness temperature of Sgr A* had been increased at the flaring event.

The flaring event of Sgr A* in 2007 occurred within one month, which is less than the typical reflective time-scale of interstellar scattering at 43 GHz. Moreover, the correlation between spectral index and flux densities reported in \citet{Lu2011} cannot be explained from simple models of interstellar scattering. Thus, the flaring event is likely to be intrinsic. Considering the features of the observed images and the cooling time-scale of electrons, it is unlikely that the flaring event is associated with an ejection of relativistic component or a temporal one-shot plasma heating such as an expanding plasma blob or a hotspot orbiting around the central black hole. Thus, the flaring event is likely to be associated with a brightness increase of the photosphere. Following self-similar ADAF \citep{Mahadevan1997}, our results do not favor the change of thermal electron temperature as the origin.  To explain the flaring event, it needs a mechanism of heating electrons continuously for much longer than orbital periods of accretion disk such as a standing shock in an accretion flow. In future, simultaneous measurements of detailed radio spectra varying from cm to sub-mm wavelengths would be helpful to constrain the property of electron distribution at the flaring event.

%---------------------------------------------------------------------------------------
% References
%---------------------------------------------------------------------------------------

\end{document}